\documentclass[prl,showpacs,preprintnumbers,amsmath,amssymb,
  nofootinbib]{revtex4}
\usepackage{amsmath}
\usepackage{amssymb}
\usepackage{amsfonts}
\usepackage{eucal}

\newcommand{\be}{\begin{equation}}
\newcommand{\bse}{\begin{subequations}}
\newcommand{\ese}{\end{subequations}}
\newcommand{\bea}{\begin{eqnarray}}
\newcommand{\eea}{\end{eqnarray}}
\newcommand{\ba}{\begin{array}}
\newcommand{\ea}{\end{array}}
\newcommand{\ee}{\end{equation}}

\begin{document}

\title{Comment on "Pulsar kicks via spin-1 color superconductivity"}

\author{Alexander Kusenko}

\affiliation{
Department of Physics and Astronomy, University of California, Los Angeles,
  CA 90024-1547, USA
}

\preprint{UCLA/05/TEP/25}

\date{\today}

\maketitle

In a recent Letter~\cite{Schmitt:2005ee}, Schmitt, Shovkovy, and Wang have
proposed a new mechanism to explain the observed velocities of pulsars. 
Unfortunately, the mechanism is not viable.

The proposed explanation is based on anisotropic emission of neutrinos
from a cooling neutron star at temperatures as low as $T \lesssim 0.1$~MeV,
some $10^3-10^6$~yr after the supernova explosion.  The anisotropy is
supposed to develop only after the nuclear matter undergoes a phase transition
to color superconductivity, which happens at a temperature well below 0.1~MeV,
tens of thousands years after the supernova explosion. 

However, the neutrino emission is negligible after the first minute since the 
onset of the supernova.  Most neutrinos are emitted during the first 10 -- 15
seconds, and almost none are produced at times after 50 seconds, when the
neutrino charged current mean free path becomes longer than the size of the
neutron star~\cite{reviews}.  When a neutron star temperature is as low as
0.1~MeV, its entire thermal energy $E_{_T}\sim 10^{48}$erg is smaller than the
kinetic energy $E_{_K}\sim 10^{49}$erg of a pulsar moving with velocity
$10^3$km/s. Therefore, in any case, the pulsar kicks must originate at some
earlier times. 

One could ask, however, whether the proposed pulsar kick mechanism could
possibly work if the phase transition happened at an earlier time, while the
neutron star is emitting neutrinos. Unfortunately, the answer is no, because
there is an additional fatal flaw in this mechanism. Neutrinos produced in an
approximate thermal and statistical equilibrium diffuse out isotropically, even
if the production cross sections and scattering amplitudes are
anisotropic~\cite{eq}.  Neutrinos produced deep in the core of a neutron star
with some initial anisotropies, whatever the origin, quickly isotropize
through scattering, and their emission becomes isotropic. Hence, they cannot
give the pulsar a kick.  This no-go 
theorem~\cite{eq} can be avoided if some of the neutrinos are out of
equilibrium, or free-streaming. Examples include ordinary neutrinos outside one
of the neutrinospheres~\cite{ks96}, sterile neutrinos~\cite{sterile},
Majorons~\cite{Farzan:2005yp}, {\it etc}. All of these represent viable
possibilities for the origin of the pulsar kicks. In particular, a sterile
neutrino with mass is the keV range is an intriguing possibility because the
same particle could be the cosmological dark
matter~\cite{sterile,kusenko_review}.   

This work was supported in part by DOE grant DE-FG03-91ER40662, as well as 
NASA grants ATP02-0000-0151 and ATP03-0000-0057.

\end{document}